\documentclass[aps,prb,twocolumn,showpacs,amsmath,amssymb,superscriptaddress]{revtex4}
\pdfoutput=1
\usepackage{graphicx}
\usepackage{amssymb}
\usepackage{natbib}
\usepackage{calc}
\usepackage{SIunits}
\usepackage{textcomp}

\begin{document}

\title{Crystal growth and characterization of a hole-doped iron-based superconductor Ba(Fe$_{0.875}$Ti$_{0.125}$)$_2$As$_2$}

\author{Yi-Li Sun}
\affiliation{Beijing National Laboratory for Condensed Matter Physics, Institute of Physics, Chinese Academy of Sciences, Beijing 100190, China}
\affiliation{University of Chinese Academy of Sciences, Beijing 100190, China} 
\author{Ze-Zhong Li}
\affiliation{Beijing National Laboratory for Condensed Matter Physics, Institute of Physics, Chinese Academy of Sciences, Beijing 100190, China}
\affiliation{University of Chinese Academy of Sciences, Beijing 100190, China} 
\author{Yang Li}
\affiliation{Beijing National Laboratory for Condensed Matter Physics, Institute of Physics, Chinese Academy of Sciences, Beijing 100190, China}
\affiliation{University of Chinese Academy of Sciences, Beijing 100190, China} 
\author{Hong-Lin Zhou}
\affiliation{Beijing National Laboratory for Condensed Matter Physics, Institute of Physics, Chinese Academy of Sciences, Beijing 100190, China}
\affiliation{University of Chinese Academy of Sciences, Beijing 100190, China} 
\author{Amit Pokhriyal}
\affiliation{Theory and Computational Physics Section, Raja Ramanna Centre for Advanced Technology, Indore 452013, India}
\affiliation{Homi Bhabha National Institute, BARC training school complex 2nd floor, Anushakti Nagar, Mumbai 400094, India} 
\author{Haranath Ghosh}
\affiliation{Theory and Computational Physics Section, Raja Ramanna Centre for Advanced Technology, Indore 452013, India}
\affiliation{Homi Bhabha National Institute, BARC training school complex 2nd floor, Anushakti Nagar, Mumbai 400094, India} 
\author{Shi-Liang Li}
\affiliation{Beijing National Laboratory for Condensed Matter Physics, Institute of Physics, Chinese Academy of Sciences, Beijing 100190, China}
\affiliation{University of Chinese Academy of Sciences, Beijing 100190, China} 
\author{Hui-Qian Luo}
\email{hqluo@iphy.ac.cn}
\affiliation{Beijing National Laboratory for Condensed Matter Physics, Institute of Physics, Chinese Academy of Sciences, Beijing 100190, China}

\date{\today}

\begin{abstract}
We report the crystal growth of a new hole-doped iron-based superconductor Ba(Fe$_{0.875}$Ti$_{0.125}$)$_2$As$_2$ by substituting Ti on the Fe site. The crystals are accidentally obtained in trying to grow Ni doped Ba$_2$Ti$_2$Fe$_2$As$_4$O.  After annealing at 500 \textcelsius  $ $  in vacuum for one week, superconductivity is observed with zero resistance at $T_{c0} \approx 17.5$ K, and about 20\% diamagnetic volume down to 2 K. While both the small anisotropy of superconductivity and the temperature dependence of normal state resistivity are akin to the electron doped 122-type compounds, the Hall coefficient is positive and similar to the case in hole-doped Ba$_{0.9}$K$_{0.1}$Fe$_2$As$_2$. The density functional theory calculations suggest dominated hole pockets contributed by Fe/Ti 3$d$ orbitals. Therefore, the Ba(Fe$_{1-x}$Ti$_{x}$)$_2$As$_2$ system provides a new platform to study the superconductivity with hole doping on the Fe site of iron-based superconductors.
\end{abstract}


\pacs{74.25.-q, 74.25.Dw, 74.70.-b, 74.25.Fy}
\maketitle

\section{Introduction}

Unconventional superconductivity usually emerges in the chemical doped or physical pressurized transition-metal-based compounds~\cite{xgwen2006,xhchen2014,kamihara2008,hdcjohnston2010,bdwhite2016,qgu2022,wwu2014,ptyang2022,mwang2024,hqluo2024}. While either external pressure or internal strain can induce superconductivity in heavy fermions, Cr-based, Mn-based and Ni-based compounds~\cite{zengwj2025,zmo2024,ostockert2024,wwu2010,jgcheng2015,hsun2023,jyli2024,ychen2025,gzhou2025,qqin2024}, chemical doping or substitution is another common way to introduce high temperature superconductivity in Cu-based and Fe-based compounds~\cite{jttranquada2014,xjzhou2021,pdai2015,grstewart2011,jswen2011}. By doping holes or electrons to suppress the long-range antiferromagnetic (AF) order, cuprates show asymmetric phase diagrams with very different critical temperature ($T_c$) and doping range~\cite{xgwen2006,jttranquada2014,xjzhou2021}. Similar case presents in the iron-based superconductors (FeSCs), but shows more flexibilities on the doped sites ~\cite{pdai2015,grstewart2011,jswen2011,rfernandes2022,hhosono2018}. Taking the typical compound BaFe$_2$As$_2$ (122-type) as an example ~\cite{mrotter2008a,ckrellner2008,qhuang2008,dlgong2017,dlgong2018}, superconductivity can be induced by hole doping on the Ba site substituted by K, Na~\cite{mrotter2008b,mrotter2008c,hqluo2008,bohmer2015,avci2014}, or electron doping on the Fe site substituted by Co, Ni, Rh, Pd, Pt, etc.~\cite{sefat2008,li2009,ni2010,canfield2009,ni2009,fhan2009,kirshenbaum2010}, or isovalent doping on the As site substituted by P ~\cite{jiang2009,kasahara2010,dhu2015,tshibauchi2014}. Such character generates different maximum $T_c$ and phase diagram even starting from the same parent compound (e.g. $T_{c, max}=39$ K for Ba$_{0.6}$K$_{0.4}$Fe$_2$As$_2$, $T_{c, max}=20$ K for BaFe$_{0.9}$Ni$_{0.1}$As$_2$, and $T_{c, max}=30$ K for BaFe$_2$(As$_{0.7}$P$_{0.3}$)$_2$), providing diverse platforms for the mechanism study of unconventional superconductivity~\cite{dhu2015,tshibauchi2014,hqluo2012,mgkim2012,xylu2013,wlzhang2016}. Since the chemical dopants not only change the charge carrier density, but also introduce disorders and local structural distortions on different atomic sites~\cite{lwang2016,hlzhou2022,wshong2023,pcheng2010,pcheng2013,jli2012}. Such side effects make a great challenge to reveal the true tuning parameters on the iron-based superconductivity, especially for the study on the electron-hole asymmetry~\cite{pdai2015,grstewart2011}.

So far as we know, superconductivity has not been achieved by hole-doping at the Fe site in FeSCs. According to previous reports in Ba(Fe$_{1-x}$\textit{TM}$_{x}$)$_2$As$_2$ (where \textit{TM} = Cr, Mn, and V), bulk superconductivity is actually absent ~\cite{ni2010,canfield2009,ni2009,mgkim2012,jli2012,wywang2017,assefat2009,yhgu2019,htakeda2014,mgkim2015}. Although the Cr or Mn substitutions on the Fe site probably introduce additional hole-type carriers, the strong local moments of Cr/Mn ions either quickly suppress the superconductivity or drive the system to a G-type magnetic state instead of the stripe-type AF order in BaFe$_2$As$_2$ ~\cite{rzhang2014,rzhang2015,dlgong2018b,dlgong2022,kmarty2011,mgkim2010,qzou2021,athaler2011,dsinosov2013,gstucker2012,mrcantarino2024}. In the V-doped BaFe$_2$As$_2$, the stripe-type AF order is gradually suppressed without the emergence of superconductivity~\cite{xgli2018}. Apart from doping on Fe site, La$_{0.5-x}$Na$_{0.5+x}$Fe$_2$As$_2$ is a rare case to explore the electron-hole asymmetry without disturbing the FeAs layer, where both types of carriers could be introduced by changing the ratio of La and Na~\cite{yhgu2018}. Unfortunately, the growth of high quality single crystals of La$_{0.5-x}$Na$_{0.5+x}$Fe$_2$As$_2$ with different doping levels is still a challenge~\cite{jqyan2015}.

Here in this work, we demonstrate that hole-type superconductivity can be realized in Ti-doped BaFe$_2$As$_2$. Ba(Fe$_{1-x}$Ti$_{x}$)$_2$As$_2$ single crystals are accidentally obtained in trying to grow Ni doped Ba$_2$Ti$_2$Fe$_2$As$_4$O~\cite{ylsun2012,mzhafiez2018}. Bulk superconductivity with $T_c \sim 21.3$ K emerges after annealing one week in the $x=0.125$ sample, where the Ni does not appear in the annealed crystals, and Ti is doped on the Fe site. The estimated anisotropy of upper critical field ($\gamma$) is about 2.0, same as other cases in the Ba-122 systems ~\cite{hqyuan2009,zswang2015}. The resistivity both in the as-grown non-superconducting and post-annealed superconducting compounds show an upward curvature and a kink around $T=50$ K, similar to the electron-type Ni or Co doped BaFe$_2$As$_2$ compounds. However, its Hall coefficient is positive and close to the value in hole-type Ba$_{0.9}$K$_{0.1}$Fe$_2$As$_2$, suggesting hole dominated charge carriers in Ba(Fe$_{1-x}$Ti$_{x}$)$_2$As$_2$, which is further confirmed by density functional theory (DFT) calculations.

\section{Experiment details}

As mentioned previously, Ba(Fe$_{1-x}$Ti$_{x}$)$_2$As$_2$ single crystals were accidentally synthesized during the attempt to grow Ni-doped Ba$_{2}$Ti$_{2}$Fe$_{2}$As$_{4}$O, which is an iron-based compound with alternative stacking along the $c-$axis with both BaFe$_2$As$_2$ and BaTi$_2$As$_2$O blocks ~\cite{ylsun2013,aablimit2017,cchsieh2018a,cchsieh2018b}. The growth procedure was actually similar to the case in the parent and Cr-doped Ba$_{2}$Ti$_{2}$Fe$_{2}$As$_{4}$O crystals by a Ba$_2$As$_3$ self-flux method. The raw materials were Ba (rods, $99+\%$), Ti (dehydrided, $>99.99\%$), Fe (powders, $99+\%$), Ni (powders, $99.99\%$), As (pieces, $99.9999\%$) and TiO$_2$ (powder, $99.9\%$). Firstly, the Ba$_2$As$_3$ flux and NiAs powder were prepared by reacting Ba pieces, Ni powders and As powders in an evacuated quartz tube.  They were heated slowly to 700 \textcelsius $ $, kept for 24 h at least, and cooled down to room temperature. Then the starting materials was mixed with a molar ratio of TiO$_2$ : Ti : Fe : NiAs : Ba$_2$As$_3$ = $1 : 3 : 3.95 : 0.05 : 6$, and pressed into pellets. The pellets were placed into an alumina crucible and sealed in an evacuated quartz tube. All operations were carried out in a glove box by keeping the water and oxygen contents less than $0.1$ ppm. The mixture was firstly heated to 1200 \textcelsius $ $ at the rate of 41 \textcelsius/h $ $ in a box furnace and kept for 24 h for completely melting down, then slowly cooled down to 850 \textcelsius $ $ at the rate of 2 \textcelsius/h. Finally, the furnace was powered off and cooled down to room temperature. By crushing the quartz tube and leaving the final products in the air for a period of time, plate-like crystals of Ba(Fe$_{0.875}$Ti$_{0.125}$)$_2$As$_2$ can be easily cleaved by separating from the powder-like flux. To obtain the superconductivity, the as-grown crystals should be annealed for about one week at 500 \textcelsius $ $ under vacuum. The self-flux method is commonly used for growing Ni and Co doped Ba-122 systems~\cite{sefat2008,li2009,ni2010,canfield2009,ni2009}, where Ni or Co can be mixed with Fe and As in the starting materials of flux. Unfortunately, we have tried to grow Ba(Fe$_{1-x}$Ti$_{x}$)$_2$As$_2$ by using FeAs and TiAs as flux, but only BaFe$_2$As$_2$ crystals can be obtained. It seems that Ti is not easily doped on the Fe site by such method, probably due to the different melting points of TiAs and FeAs.

Single-crystal X-ray diffraction (XRD) measurements were carried out on a Bruker D8 VENTURE diffractometer equipped with a Mo~K$\alpha$ radiation source ($\lambda = 0.7107$~\AA) at room temperature. The crystalline quality and lattice parameters were analyzed from the diffraction data. The crystal surface morphology and element contents were checked by energy dispersive X-ray (EDX) with S-4800 Scannning Electron Microscope (SEM). The sample composition was preliminary determined as the average results from the analysis of the EDX spectrum on random areas of several single crystals, and further checked by the inductively coupled plasma (ICP) analysis. The Laue pattern was collected by an X-ray Laue camera (Photonic Sciences) in backscattering mode with the incident beam along the $c-$axis. The out-of-plane diffraction was measured using a 9~kW high-resolution single-crystal X-ray diffractometer (SmartLab) with Cu~K$\alpha$ radiation ($\lambda = 1.5406$~\AA) at room temperature in reflection mode, with the $2\theta$ range from $10^\circ$ to $80^\circ$. The resistivity and Hall measurements were carried out on a Quantum Design physical property measurement system (PPMS) by the standard four-probe or six-probe method, respectively. The DC magnetization was measured with the zero-field-cooling (ZFC) method under a small magnetic field $H$ = 10 Oe perpendicular to $c-$axis on a Quantum Design Magnetic Property Measurement System-3 (MPMS-3).

The electronic structures are calculated using DFT method with the plane wave basis set approach implemented in Quantum ESPRESSO package ~\cite{giannozzi2017}, employing the Perdew-Burke-Ernzerhof (PBE) form of the general gradient approximation for the exchange-correlation functional ~\cite{perdew1996}. The calculations utilized the experimental crystal structures with space group $I4/mmm$ (No. 139), lattice constants $a = 3.9834 \, \AA$, $c = 13.193 \, \AA$, and $z_{As}= 0.3559$. Ti-doping was simulated using the virtual-crystal approximation, where the Fe sites are occupied by 12.5\% of Ti with same valence states by generating a new VCA pseudopotential interpolating between the two existing pseudopotentials of Fe and Ti.~\cite{bellaiche2000}. The Monkhorst-Pack scheme was used for Brillouin zone sampling in $\mathit{k}$-space ~\cite{monkhorst1976, pack1977}. For self-consistent field calculations, a $\mathit{\Gamma}$-centered ($16 \times 16 \times 16$) $\mathit{k}$-mesh was used, with a ($20 \times 20 \times 20$) $\mathit{k}$-mesh for non-self-consistent field calculations. Ultrasoft pseudopotentials from the Psilibrary database are used ~\cite{dal2014}, and the Kohn-Sham orbitals are expanded in a plane wave basis set with a kinetic energy cutoff of $70$ $ \mathit{R_{y}}$ for the Ti-doped compound. Gaussian smearing is applied with a smearing parameter of $0.005$ $\mathit{R_{y}}$, and the charge density cutoff is set to ten times the plane wave energy cutoff. The energy convergence threshold for scf calculations was $10^{-8}$ (a.u). A $40 \times 40 \times 40$ $\mathit{k}$-mesh was employed to converge the Fermi surface and density of states (DOS), with a Gaussian smearing parameter of $0.002$ $\mathit{R_{y}}$. The visualization of Fermi surface and Fermi velocity was performed using the FermiSurfer visualization tool ~\cite{kawamura2019}. In all calculations, the $z-$direction is defined as perpendicular to the Fe-Fe square lattice, while the $x$ and $y$ directions are aligned with the nearest Fe-Fe bond directions.

\section{Results and discussion}

\begin{figure}[t]
\includegraphics[width=0.46\textwidth]{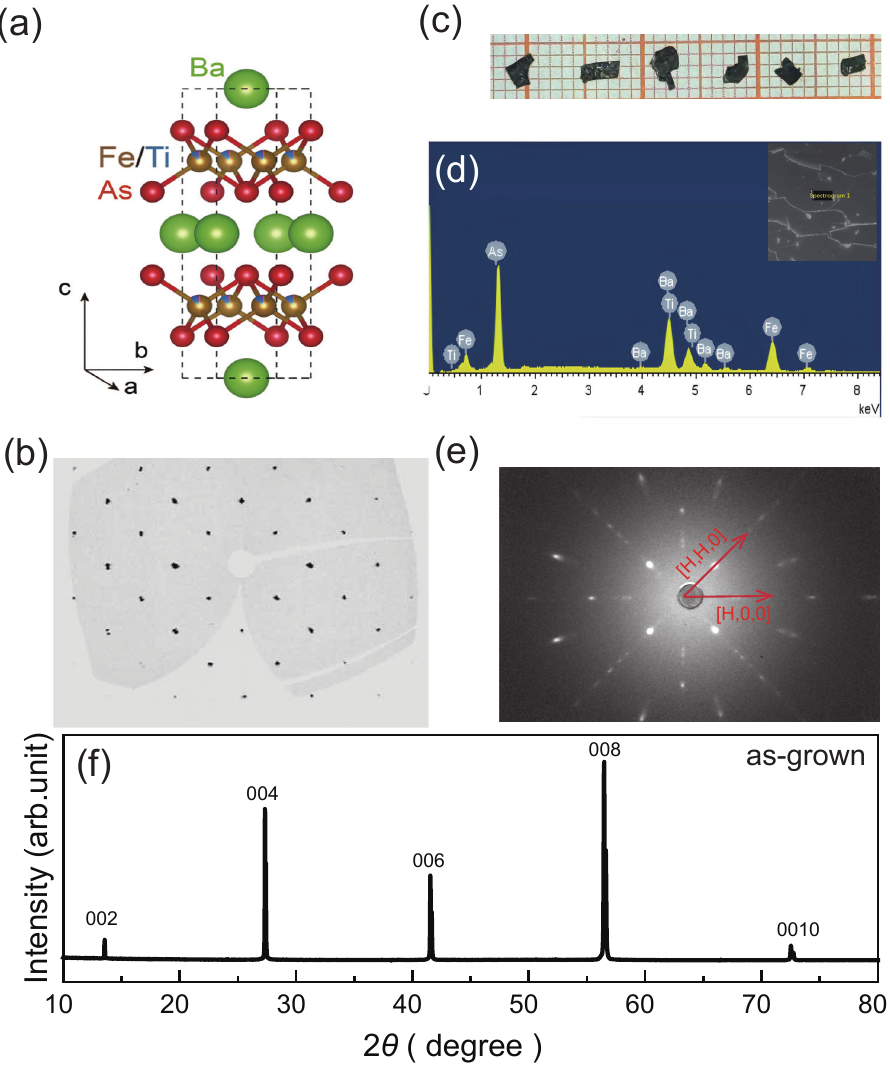}\\[5pt]
\caption{(a) Crystal structure of Ba(Fe$_{0.875}$Ti$_{0.125}$)$_{2}$As$_{2}$. (b) XRD pattern of one typical single crystal. (c) Photo of as-grown crystals. (d) Typical EDX spectrum on one crystal. The inset is the SEM photograph of this crystal, showing a flat surface and a layered structure. (e) Typical Laue reflection pattern for our crystals. (f) Out-of-plane XRD with (0, 0, $L$) ($L=$ even) reflections at room temperature for one as-grown single crystal.}	
\end{figure}

\begin{table}
	{\footnotesize{\bf Table 1.} Results of the chemical composition determined by ICP and EDX.  Here, the detection limit is about 1~\% for ICP and 2~\% for EDX, respectively, but the obtained results are averaged values from several measurements with statistical error $\sim$5~\%. \\
		\vspace{2mm}
		\begin{tabular}{cccccccc}
			\hline
			&Element &Ba&Fe&Ti&As&Ni&O\\
			\hline
			&ICP atomic ratio (\%) & 20.03 & 34.90&5.18&39.44&0&0\\
			\hline
            &EDX atomic ratio (\%) & 24.13 & 35.58&6.20&34.08&0&0\\
			\hline
	\end{tabular}}
\end{table}

By attempting to substitute Ni into Ba$_2$Ti$_2$Fe$_2$As$_4$O, we have finally obtained Ba(Fe$_{1-x}$Ti$_{x}$)$_2$As$_2$ single crystals instead. After annealing for one week under vacuum, the crystals become superconducting. Figure 1 shows the results of crystalline quality of our crystals. Notably, neither the single crystal XRD analysis on the bulk samples nor the EDX analysis on the freshly cleaved surfaces could detect the existence of Ni. The actual concentration for each element determined by the ICP method (Table 1) with a detection limit about 1{\%} give a chemical formula: Ba(Fe$_{0.875}$Ti$_{0.125}$)$_2$As$_2$, where the statistics errors are about 5{\%} from the analyses on several pieces of crystals. Again, no Ni is observed in the superconducting sample, thus it rules out the possibility of Ni doped BaFe$_2$As$_2$ in our crystals. We believe that Ni probably acts as the catalyst during the growth of Ti-doped BaFe$_2$As$_2$, which may be helpful to stabilize the crystal structure in analogy to the 112-type FeSC ~\cite{sjiang2016,xie2018}. It seems very difficult to grow Ba(Fe$_{1-x}$Ti$_{x}$)$_2$As$_2$ by directly using the starting materials of Ba, FeAs and TiAs, where BaFe$_2$As$_2$ is usually obtained. The single crystal XRD suggests an isostructure of BaFe$_2$As$_2$ with I/4mmm space group, namely, the 122-type of FeSC ~\cite{mrotter2008a,mrotter2008b,mrotter2008c}. Although our single-crystal data lack sufficient quality for definitive refinement on the atomic coordinates or bond angles, it clearly suggests that Ti atoms are substituted on Fe sites. The lattice parameters are calculated to be $a=3.9834$ ${\rm \AA}$ and $c=13.193$ ${\rm \AA}$ [Fig.1(a) and Fig.1(b)], where the in-plane lattice parameter is larger than BaFe$_2$As$_2$ ($a=3.9625$ ${\rm \AA}$) due to the Ti substitution effect ~\cite{xhchen2014,grstewart2011,mrotter2008a}. The photo in Fig. 1(c) shows the typical sizes of our Ba(Fe$_{0.875}$Ti$_{0.125}$)$_2$As$_2$ crystals. The dimensions of the largest crystal are about 4 mm $\times$ 3 mm $\times$ 0.5 mm. The cleaved surface is shiny under the light, and the texture is brittle. Figure 1(d) displays a typical EDX spectrum of our crystals with four elements (Ba, Fe, Ti, As) detected with a detection limit about 2{\%}. The element distribution is homogenous on the cleaved surface. By averaging the EDX results on several random area of the crystal, we estimate the composition is Ba(Fe$_{0.737}$Ti$_{0.128}$)$_2$As$_{1.4}$, slightly different from the ICP results (Table 1). The Laue reflection pattern is shown in Fig. 1(e). Again, the high quality of our sample results in bright and sharp scattering spots, clear orientations along [$H$, 0, 0] and [$H$, $H$, 0] in the reciprocal space can be observed. The out-of-plane XRD with (0, 0 ,$L$) reflections for one as-grown crystal is shown in Fig. 1(f). Only $L=$ even peaks can be observed, suggesting the centrosymmetric (ThCr$_2$Si$_2$-type, space group $I4/mmm$) structure  similar to BaFe$_2$As$_2$ ~\cite{grstewart2011,mrotter2008a,mrotter2008b,mrotter2008c}. The $c$-axis lattice constant is determined to be $13.190~\text{\AA}$, which is in good agreement with the single crystal XRD analysis, but slightly larger than the $c-$axis in BaFe$_2$As$_2$ ($c=13.017~\text{\AA}$) ~\cite{mrotter2008a}.

Figure 2(a) shows the temperature dependence of the in-plane longitudinal electrical resistance ($\rho_{xx}(T)$) under zero magnetic field. The data is normalized by the resistance at $300$ K for comparison. Both as-grown and annealed samples show a metallic behavior when initially cooling down from room temperature, and a kink follow by slight upturn emerges around $T=50$ K. Such response in resistance could be a signature of magnetic transition, which is common in 122-type FeSCs. After annealing, the sample becomes a bit more metallic along with a superconducting transition with onset temperature at $T_{c,onset}=21.3$ K and zero resistance temperature at $T_{c0}=17.5$ K. In fact, the normal state resistance with an upward curvature behaves like the underdoped BaFe$_{2-x}$Ni$_x$As$_2$ ~\cite{bshen2011,nni2008,ychen2011}, which is a typical electron-type FeSC. Figure 2(b) shows the magnetization results of the annealed sample under a magnetic field of $H=10$ Oe in $ab-$plane ($H \parallel ab$). The distinct difference between FC and ZFC data suggests the sample is a type-II superconductor. The magnetic susceptibility $4\pi\chi \approx -0.2$ at 2 K indicate 20\% diamagnetic volume, namely bulk superconductivity but not full Meissner state in Ba(Fe$_{0.875}$Ti$_{0.125}$)$_2$As$_2$. However, the onset temperature of the dropdown in ZFC magnetization is 15.5 K, slightly lower than the zero resistance temperature [inset of Figure 2(b)], suggest possible weak links between the superconducting islands inside the sample.  Indeed, the magnetization transition in Fig.2(b) is very broad, this could be attributed to the inhomogeneity or disorders in the annealed sample, where both non-superconducting and superconducting regions intertwine with each other. Therefore, the annealing condition could be further optimized. For instance, annealing the 122-type FeSCs in Ba$_2$As$_3$ will effectively reduce the disorders \cite{kasahara2010,dhu2015,tshibauchi2014}. Figure 2(c) shows the field dependence of the Hall (transverse) resistivity $\rho_{xy}$ at different temperatures for $H \parallel c$, which is measured by sweeping both the negative and positive magnetic fields at several fixed temperatures, to eliminate the effects from asymmetric geometry of the electric contacts [$\rho_{xy}=(R_{xy}(+)-R_{xy}(-))/2$]. In the present temperature region,  all $\rho_{xy}$ show linear field dependences with positive slopes, confirming that the Ba(Fe$_{0.875}$Ti$_{0.125}$)$_2$As$_2$ mainly hosts hole-type charge carriers. Hereby, we deduce the temperature dependence of the Hall coefficient $R_H$ from $\rho_{xy}$, as shown in Figure 2(d). The sign of $R_H$ is positive in the whole temperature range and continues to decrease upon warming up to 300 K. Just above $T_c$, the absolute value of $R_H$ is seven times larger than the case at 300 K. Such behaviors indicate that Ba(Fe$_{0.875}$Ti$_{0.125}$)$_2$As$_2$ is a multi-band system similar to the hole-doped system such as Ba$_{1-x}$K$_{x}$Fe$_{2}$As$_{2}$ ~\cite{bshen2011}. Meanwhile, it probably reveals that the mobility of the hole induced by Ti-doping becomes the dominant factor as temperature decreases. We also notice that the absolute $R_H$ in Ba(Fe$_{0.875}$Ti$_{0.125}$)$_2$As$_2$ has a similar magnitude with Ba$_{0.9}$K$_{0.1}$Fe$_2$As$_2$ ~\cite{hqluo2009}.

\begin{figure}[t]
\includegraphics[width=0.5\textwidth]{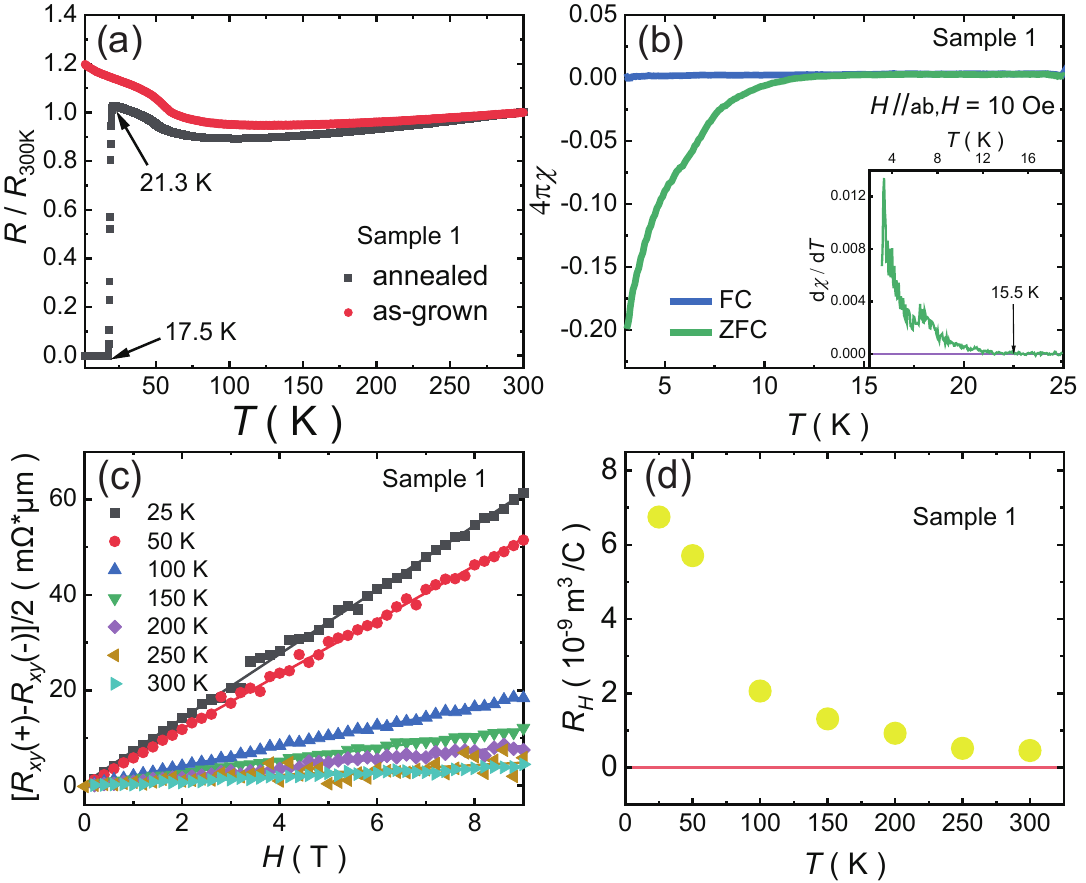}\\[5pt]
\caption{(a) The temperature dependence of resistance for as-grown and annealed Ba(Fe$_{0.875}$Ti$_{0.125}$)$_2$As$_2$ sample 1. All data are normalized by the room temperature resistance. (b) Temperature dependence of the DC-magnetic susceptibility measured by field-cooling (FC) and zero-field-cooling (ZFC) methods under a small field $H=10$ Oe in $ab-$plane ($H \parallel ab$). The inset is the first-order derivative of the ZFC data. (c) Field dependence of Hall resistivity $\rho_{xy}$ at various temperatures with $H \parallel c$. (d) Temperature dependence of the Hall coefficient $R_H$.}
\end{figure}
		
\begin{figure}[t]
\includegraphics[width=0.5\textwidth]{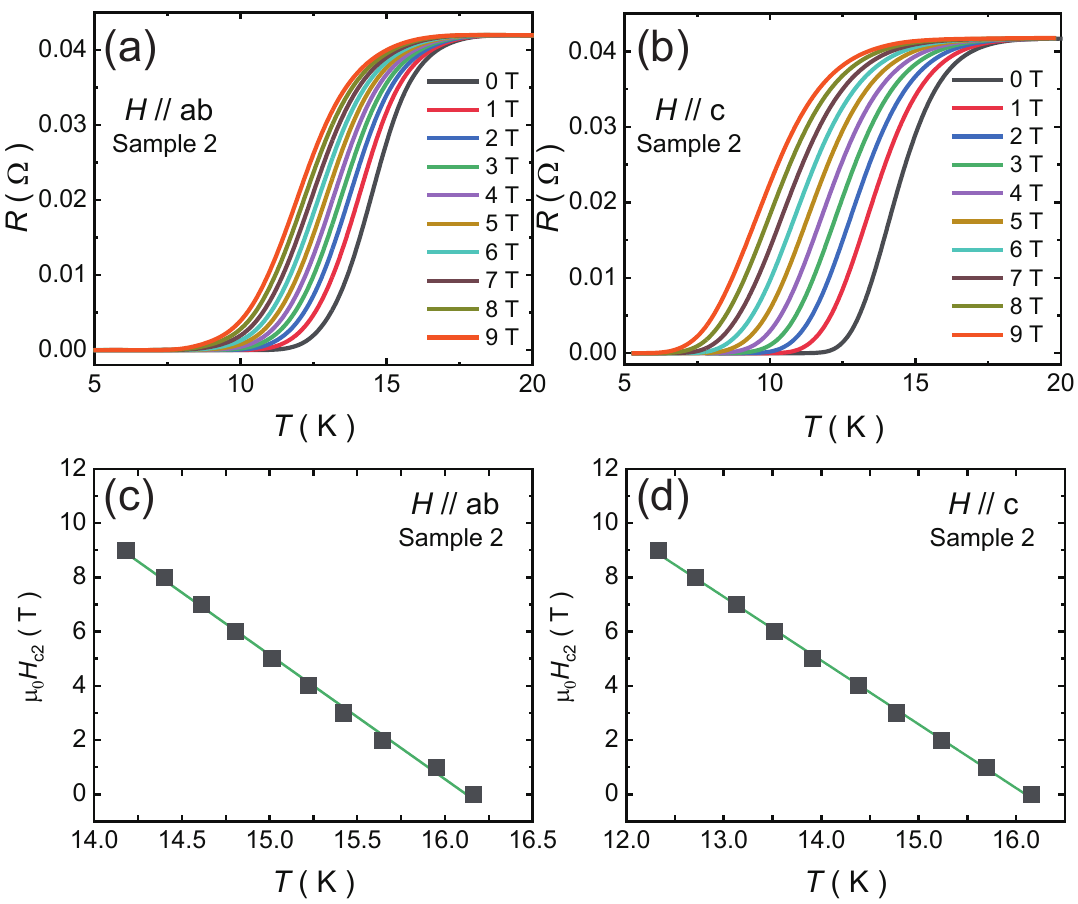}\\[5pt]
\caption{(a) and (b) Suppression of the superconductivity of sample 2 under magnetic fields for $H \parallel ab$ and $H \parallel c$, respectively. (c) and (d) The upper critical field $H_{c2}$ obtained from the results in (a) and (c). The red straight lines are linear fittings to the results closed to $T_c$ for the estimation of their slopes. }
\end{figure}

We choose another crystal as Sample 2 ($T_{c,onset}=17$ K and $T_{c0}=12$ K) to measure the magnetic field suppression on superconductivity. The results are presented in Fig. 3 both for $H \parallel ab$ and $H \parallel c$. As the field increases, the superconducting transition shifts to low temperatures, indicating weak flux pinning effect in this material. The suppression of $T_c$ for the field along the $c-$axis is more serious than that for $H \parallel ab$ due to the anisotropy of superconductivity.  Additionally, we plot the upper critical field $H_{c2}$ within the $ab$ plane and along the $c$-axis in Fig. 3 (c) and (d), respectively, which is determined by the temperature function of the 90\% resistance value during the superconducting transition. The slopes of $H_{c2}$ along $ab$ - and $c$ - direction at $T_c$ are -4.5 T/K and -2.3 T/K, respectively. According to the Werthamer-Helfand-Hohenberg theory~\cite{werthamer1966,zswang2015}, we roughly estimate  $H_{c2(0)}$ = 66.1 T and 33.8 T along $ab$ - and $c$ - direction, respectively. Then the anisotropy of the upper critical field $\gamma$ = $H\substack{^{ab}_{c2}}$ / $H\substack{^c_{c2}} {\approx} 2$, much similar to other 122-type FeSCs ~\cite{hqyuan2009,zswang2015,nni2008}.

Finally, Fig. 4 displays the results of DFT calculations in Ba(Fe$_{0.875}$Ti$_{0.125}$)$_2$As$_2$ using the same lattice constants obtained from the single crystal XRD results. Figure 4(a) shows the atom-projected partial density of states. The pattern indicates the states near the Fermi level are predominantly contributed by Fe/Ti, with a smaller contribution from As, and no charge transfer is indicated in between Fe and Ti. Figure 4(b) shows the band structure with orbital projections. Bands crossing the Fermi level are  primarily, $d_{xz}$, $d_{yz}$, and $d_{xy}$ orbitals derived. Notably, three hole-like bands are present near the $\Gamma$-point, while two electron-like band appears around the high-symmetry $M$-point. The strongly dispersive innermost hole band near the $\Gamma$ point is derived from the $d_{xz}$ orbital, while out of two other hole-like bands, one is $d_{xy}$ derived and the other is $d_{yz}$ derived with some contribution of $d_{x^2-y^2}$ character. At high symmetry $M$-point, the bottom of the $d_{xy}$ + $d_{xz/yz}$ derived electron-like band is almost at the Fermi level, whereas the $d_{xy}$ electron band has its bottom edge well below the Fermi level. Furthermore, while most of the bands are flat along the $\Gamma$-$Z$ direction, two bands with predominant $d_{z^2}$ character exhibit strong dispersion. Figure 4(c) and (d) depicts the Fermi surface with color intensity indicating Fermi velocity, ranging from blue to yellow. Two visible hole-like bands are apparent, with the outer two being nearly degenerate, as shown in the band structure of Fig. 4(b). Significant change of Fermi velocity is observed for the inner hole pockets along the $\Gamma$-$Z$ direction, suggesting possible three-dimensional behaviors in consistent with the low superconducting anisotropy. A relatively small electron-like pocket is found at around $M$ point, which could potentially vanish with minor modifications, such as doping or pressure induced possible Lifshitz transition. Overall, the Fermi surfaces of Ba(Fe$_{0.875}$Ti$_{0.125}$)$_2$As$_2$ are dominated by hole pockets.

\begin{figure}[t]
\includegraphics[width=0.5\textwidth]{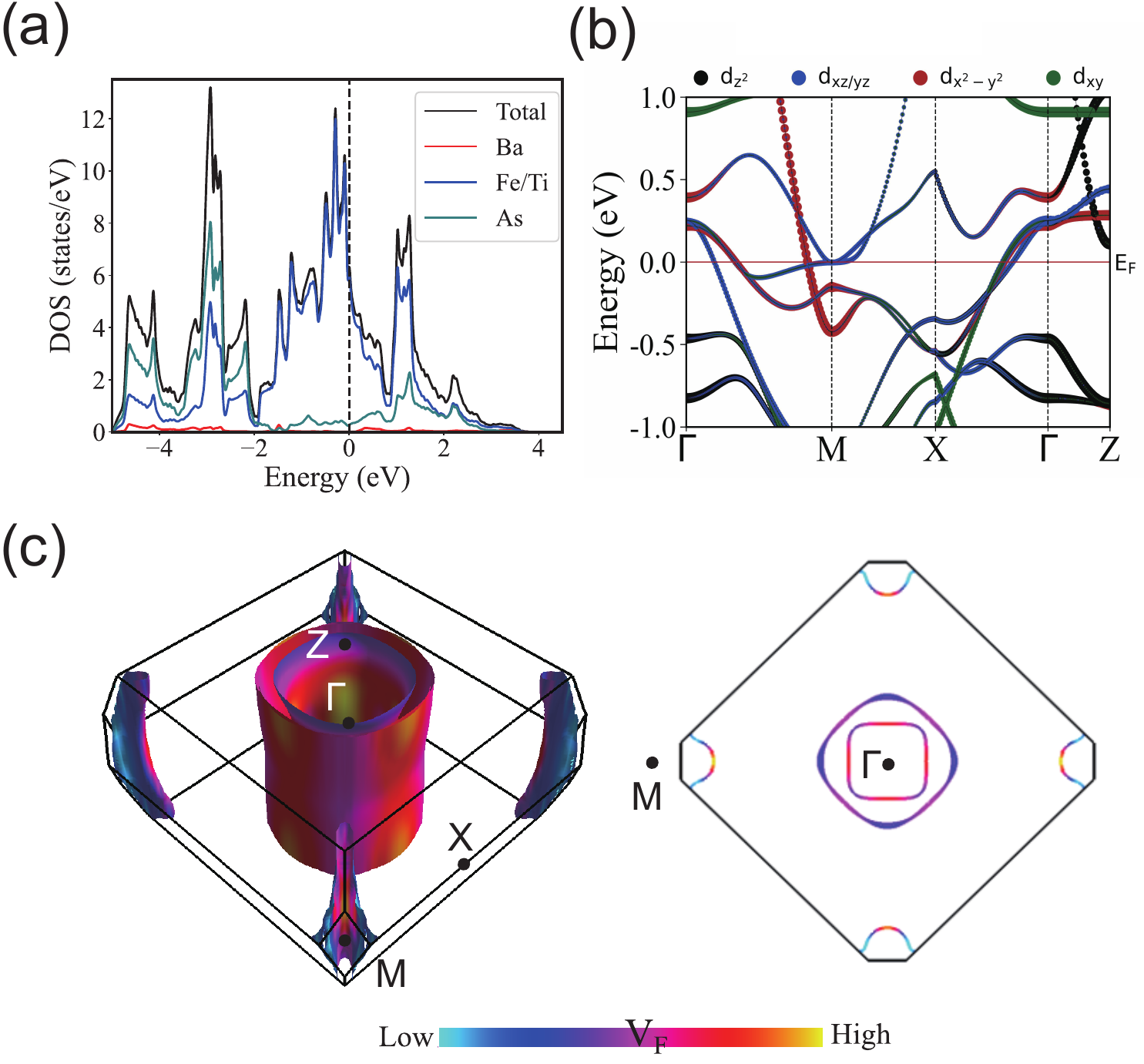}\\[5pt]
\caption{DFT calculation results of Ba(Fe$_{0.875}$Ti$_{0.125}$)$_2$As$_2$. (a) Density of states (DOS) for individual atomic species. (b) Band structure along high-symmetry directions, orbital contributions are depicted with colored circles, where the size of each circle represents the weight of the respective orbital. (c) and (d) Fermi surfaces and the corresponding Fermi velocity in the Brillouin zone, with high-symmetry points indicated for reference.}
\end{figure}

Our results provide the first example of single crystal growth of superconducting Ba(Fe$_{1-x}$Ti$_{x}$)$_2$As$_2$ ($x=0.125$) compound with hole doping on the Fe site. Although the obtained crystals are byproducts during the growth of Ni-doped Ba$_{2}$Ti$_{2}$Fe$_{2}$As$_{4}$O, the possible catalyzing effect from Ni seems helpful during the synthesis process thus inspires us to explore new FeSCs, especially for those compounds with alternative stacking structure~\cite{hjiang2013}. The realization of superconductivity in Ba(Fe$_{1-x}$Ti$_{x}$)$_2$As$_2$ provides several new clues for the study on the doping effect in FeSCs. First, it seems that the curvature of $\rho_{xx}(T)$ in the normal state does not correspond to the charge carrier type. For instance, the resistivity in the electron-type BaFe$_{2-x}$Ni$_x$As$_2$ shows an upward curvature~\cite{hqluo2012,nni2008,ychen2011}, but an downward curvature of $\rho_{xx}(T)$ is always observed in the hole-type Ba$_{1-x}$K$_{x}$Fe$_{2}$As$_{2}$~\cite{bshen2011}. Second, doping on the Fe site may accompany more disorders, thus the maximum $T_c$ is below those cases for doping on Ba or As sites. Even though we do not establish the phase diagram of the hole-type Ba(Fe$_{1-x}$Ti$_{x}$)$_2$As$_2$ system yet, its superconductivity seems more sensitive to the annealing process rather than the Ti doping level (i.e. sample 1 and sample 2 have different $T_c$). Besides the doping induced superconductivity, post-annealing is an effective way to minimize the effects from the scattering of impurities and disorders, especially for the Ba$_{2}$Ti$_{2}$Fe$_{2}$As$_{4}$O and Fe$_{1+y}$Te$_{1-x}$Se$_x$ systems ~\cite{pdai2015,grstewart2011,jswen2011,ylsun2013}. Finally, since FeSCs are multi-band and metallic systems, chemical doping or charge carrier density is not a sole scaling parameter to describe the superconductivity. We notice that even in those compounds dominated by hole pockets, such as hole-overdoped Ba$_{1-x}$K$_{x}$Fe$_{2}$As$_{2}$, KFe$_{2}$As$_{2}$ and KCa$_2$Fe$_4$As$_4$F$_2$, superconductivity with $s^{\pm}$ pairing symmetry could exist accompanied by strong spin fluctuations ~\cite{dswu2024,yli2025}.  The contrary cases with only electron pockets also show superconductivity in iron selenides~\cite{xjzhou2021,pdai2015,grstewart2011,jswen2011}.

The emerged superconductivity in Ba(Fe$_{1-x}$Ti$_{x}$)$_2$As$_2$ is possibly related to the itinerant nature of Ti, which simultaneously contributes the charge carrier to the Fermi level and preserves the correlated Hund's metal state in the parent compound ~\cite{xhchen2014,pdai2015,grstewart2011,rfernandes2022}. In principle, the chemical balance requires the Ti valence to be near +2 (Ti$^{2+}$, 3d$^2$) similar to the cases of Cr (Cr$^{2+}$, 3d$^4$) and Mn (Mn$^{2+}$, 3d$^5$) doped compounds, all of them should contribute holes for their less 3d electrons than Fe$^{2+}$ (3d$^6$) \cite{kmarty2011,mgkim2010,qzou2021,athaler2011,dsinosov2013,gstucker2012,mrcantarino2024,slafuerza2017,ytexier2012,hsuzuki2013,tkobayashi2016,mrcantarino2023,fagarcia2019,lchen2021,mrcantarino2025}. However, the dopant Mn 3d states are almost localized in spite of the strong Mn 3d-As 4p hybridization, driving the system to a Hund's insulator in BaMn$_2$As$_2$ \cite{slafuerza2017,ytexier2012,hsuzuki2013,tkobayashi2016}. Such magnetic impurity effects from Mn dopant significantly suppress the stripe-type magnetic excitations in BaFe$_2$As$_2$\cite{mgkim2010,qzou2021,athaler2011}, and introduce competing G-type antiferromagnetic order or Griffiths phase associated with N\'{e}el-type excitations\cite{dsinosov2013,gstucker2012,mrcantarino2023,fagarcia2019,lchen2021}, which do not support the $s^{\pm}$ superconductivity in FeSCs. Although the Cr substitution acts as a hole dopant at low concentration, the magnetic impurity and disorder effects dominate over the charge doping effect, leading to a softening of Fe-derived magnetic excitations primarily by damping \cite{rzhang2015,dlgong2018b,dlgong2022,kmarty2011,mrcantarino2024,mrcantarino2025}. Therefore, the absence of superconductivity in Cr and Mn-doped BaFe$_2$As$_2$ is likely linked to the joint effect of increasing correlation and disorder effects on the Fe lattice introduced by magnetic impurities \cite{mrcantarino2024,mrcantarino2025}. Instead, the non-magnetic dopants like Ti may give a chance to induce superconductivity when the charge doping overrides disorder effects and preserves the spin fluctuations. Therefore, to tune the $T_c$ and superconducting states in FeSCs, one must comprehensively consider the effects from charge doping, lattice distortion and disorder effects, all of them could affect the electronic structure and the magnetic interactions.

\section{Summary}

To summarize, we have accidentally grown a new single crystal of FeSC, Ba(Fe$_{0.875}$Ti$_{0.125}$)$_2$As$_2$, by doping Ti at the Fe site. The annealed crystals show bulk superconductivity with zero resistance at $T_{c0} \approx 17.5$ K. Both the positive Hall coefficient in transport measurements and the dominated hole pockets in DFT calculations suggest that it is a hole-type compound. Therefore, the Ba(Fe$_{1-x}$Ti$_x$)$_2$As$_2$ system may be a new platform to study the doping effects on the superconductivity in FeSCs.

\addcontentsline{toc}{chapter}{Acknowledgment}
\section*{Acknowledgment}
This work is supported by the National Key Research and Development Program of China (Grants No. 2023YFA1406100, No. 2022YFA1403800, No. 2022YFA1403400 and No. 2021YFA1400400), the National Natural Science Foundation of China (Grants No. 12274444 and No. 12574165), the Chinese Academy of Sciences (Grant No. XDB25000000).DFT calculations are performed at the HPC facilities at RRCAT. AP acknowledges financial support from HBNI-RRCAT.

\end{document}